**Submitted to Radiation Measurements**

# Hadron Fluence Measurements with LiF-TLD Sensors at the Proton Synchrotron Accelerator at CERN

Christoph Ilgner[a], Maciej Budzanowski[b], and Barbara Obryk[b]

[a]*Experimenelle Physik 5, Technische Universität Dortmund, 44221 Dortmund, Germany*

[b]*Radiation Physics and Dosimetry Department of the Institute of Nuclear Physics, Polish Academy of Sciences, ul. Radzikowskiego 152, 31-342 Kraków, Poland*

**Abstract**

In view of the implementation of beam-monitoring sensors for CERN's Large Hadron Collider (LHC), and also in order to validate Thermoluminescence Detectors as a versatile tool to measure ionizing radiation doses in mixed fields at hadron colliders such as the LHC, chemical vapor deposition diamond sensors have been evaluated and calibrated at CERN's Proton Synchrotron accelerator. Special attention was paid to understanding whether lithiumfluoride thermoluminescence detectors are suitable as measuring devices in these radiation fields.

## 1. Introduction/Scope

Particle physics experiments installed at high-luminosity hadron accelerators such as the LHC need to be able to cope with possible adverse radiation conditions. In the case of the LHC, these are particularly hadronic showers from misaligned beams hitting structure material. In such a case, beam particles would interact with matter, causing a significant build-up in ionizing radiation dose which potentially destroys sensitive detector components. More and more widely, chemical vapor deposition (CVD) diamonds are being used as the sensor material for detectors that protect the experiments from that kind of damage.

For validation purposes, such a CVD diamond sensor has been exposed to a 3.5 GeV hadron beam from CERN's Proton Synchrotron [1]. In order to measure the fluence of these hadrons, i.e. protons and pions, the diamond sensor has been replaced temporarily by a scintillator of 5 x 5 $mm^2$ for particle counting at lower fluxes.

In a second step, matrices consisting of 7 × 7 and 5 × 5 pellets of $^6$LiF thermoluminescence (TLD) sensors of a surface of 3.1 mm × 3.1 mm have been introduced into the beam at various hadron fluxes. The higher fluence values have been obtained with photographic emulsions and aluminum activation.

The performance of the $^6$LiF thermoluminescence sensors will be outlined more in detail in the following sections.

## 2. The test setup

As shown in Fig. 1, matrices of thermoluminescence detectors made from $^6$LiF were brought into the hadron beam and have been exposed to several particle fluences, with the diamond sensor in line with the TLD matrices.

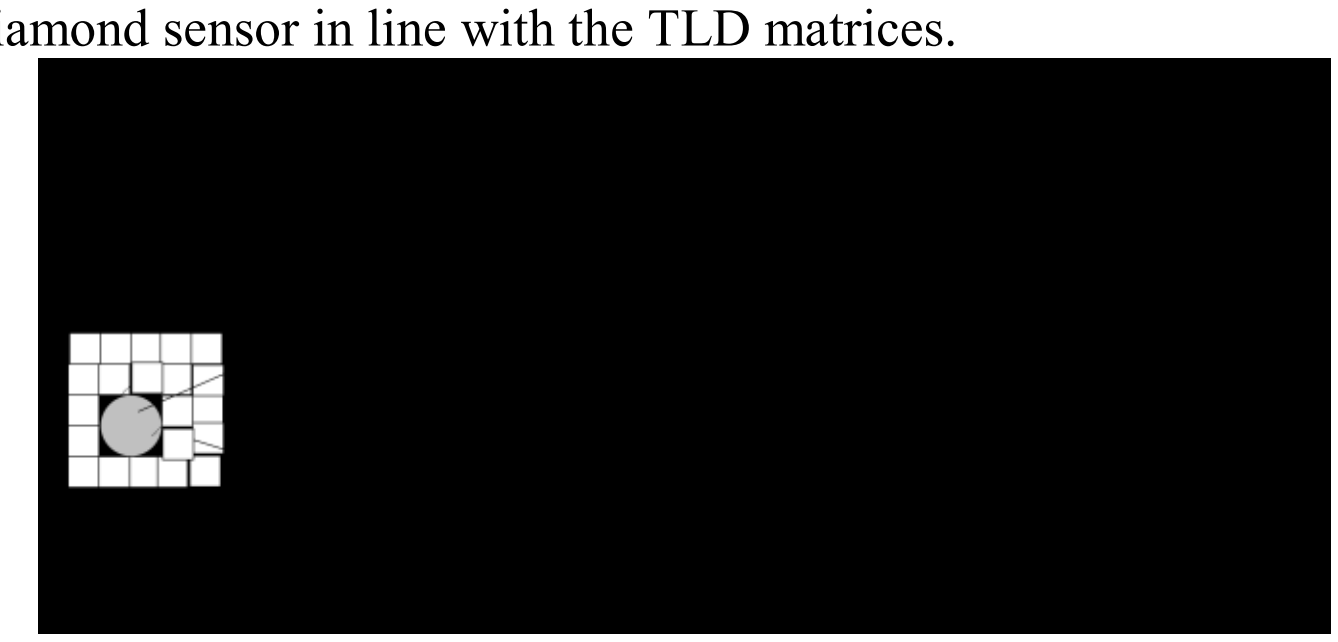

Fig. 1: The TLD matrix used to determine the profile of the hadron beam.

For calibration purposes, a scintillator of 5 mm × 5 mm has been placed behind the diamond sensor for particle counting.

For proton fluxes that saturated the scintillator, aluminum sheets were used, in which the activity of $^{22}$Na was determined after irradiation.

A typical train of proton bunches shined onto the diamond detector is shown in Fig. 2, where the current signal from that sensor is plotted.

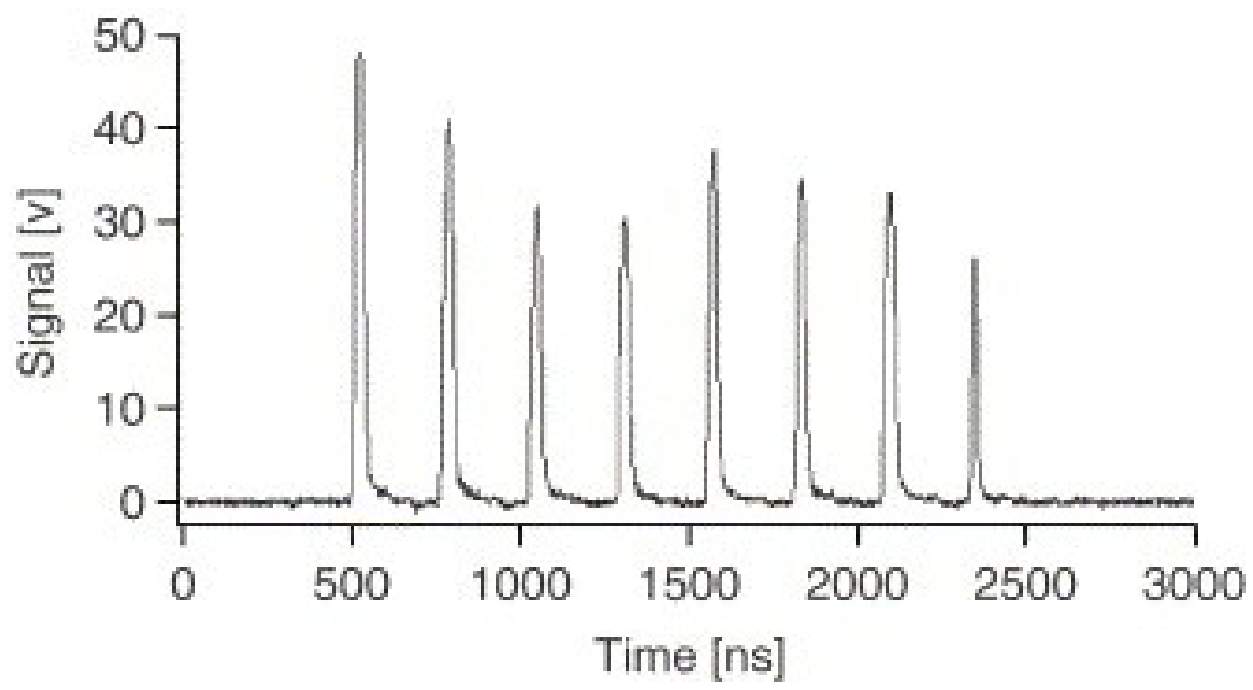

Fig. 2: Response signal of a 300 µm thick diamond sample, biased at 300 V, to a train of eight bunches of approx. $10^8$ protons/cm$^2$, separated by 262 ns [2].

## 2. Analysis of the TLD response

After exposure, the TLD have been analyzed on both the Harshaw 5500 and ALNOR Dosacus TLD reader. Since these readers give out data in different formats, based on the integrated photomultiplier current, which, for the Harshaw reader, needs to be operated at varying high voltages, calibrations have been performed to assure comparability of the measured values. The data obtained during the diamond exposure to the test beam were corrected for these readout conditions.

In order to cope with the large range of hadron fluences, which extends over four orders of magnitude, for the Harshaw device, the PMT is operated at two different gains, corresponding to high voltages of 614V (low gain) and 860V (high gain). Since the response of LiF material to intense hadron fields is not known, for this reader a second temperature profile has been applied for TLD readout as shown in Fig. 3. This appeared to be reasonable as a cross check, since with temperature profile 2 the TLD material, of which a potential radiation damage was not known initially, was kept longer at the end temperature of 300°C [3].

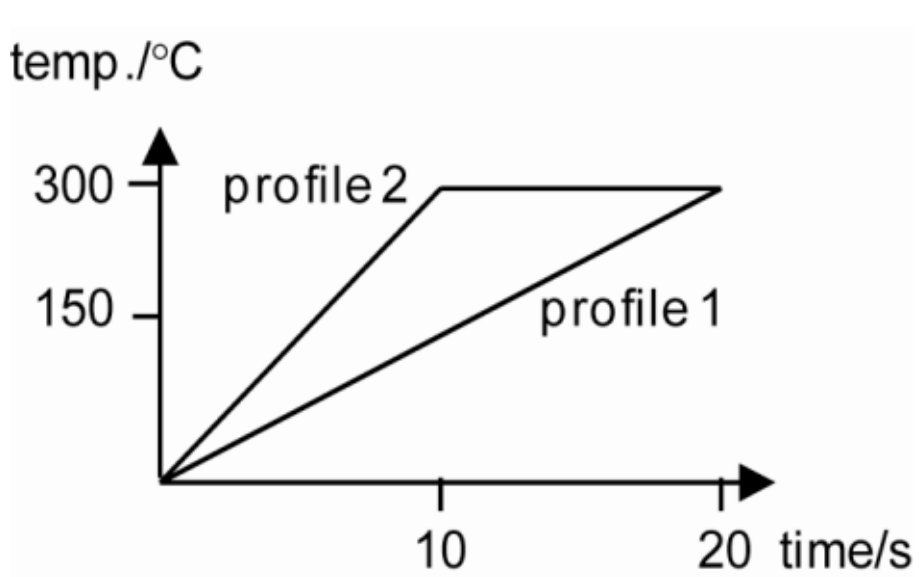

Fig. 3: Temperature profiles for TLD readout with the Harshaw reader.

These different operating regimes appeared to be necessary, since the particle fluence range covered almost four orders of magnitude. In order to assure consistency, correction factors were worked out that allowed to normalize the obtained dose values. For this, TLD pellets were irradiated with a particle fluence of $(5\pm1)\cdot10^6$ and $(18\pm1)\cdot10^6$ particles. With the dose values obtained for the different readout regimes, the standard particle fluence (dose) value $D_{LGTP1}$ at low photomultiplier gain and temperature profile 1 could be retrieved from the dose value $D_{xyz}$ obtained at that different readout regime by

$$D_{LGTP1} = c_{xyz}(D_{xyz}) \cdot D_{xyz} \,,$$

Where $c_{XYZ}$ is the correction factor for the respective readout regime. The correction factors are given in Table 1; since they depend on the dose measured, a linear fit was applied with good results as shown in Fig. 4 and 5, according to

$$c_{xyz}(D_{xyz}) = ax+b,$$

where $a$ and $b$ are the parameters for the linear fit, and $x$ is the dose equivalent in scintillator counts.

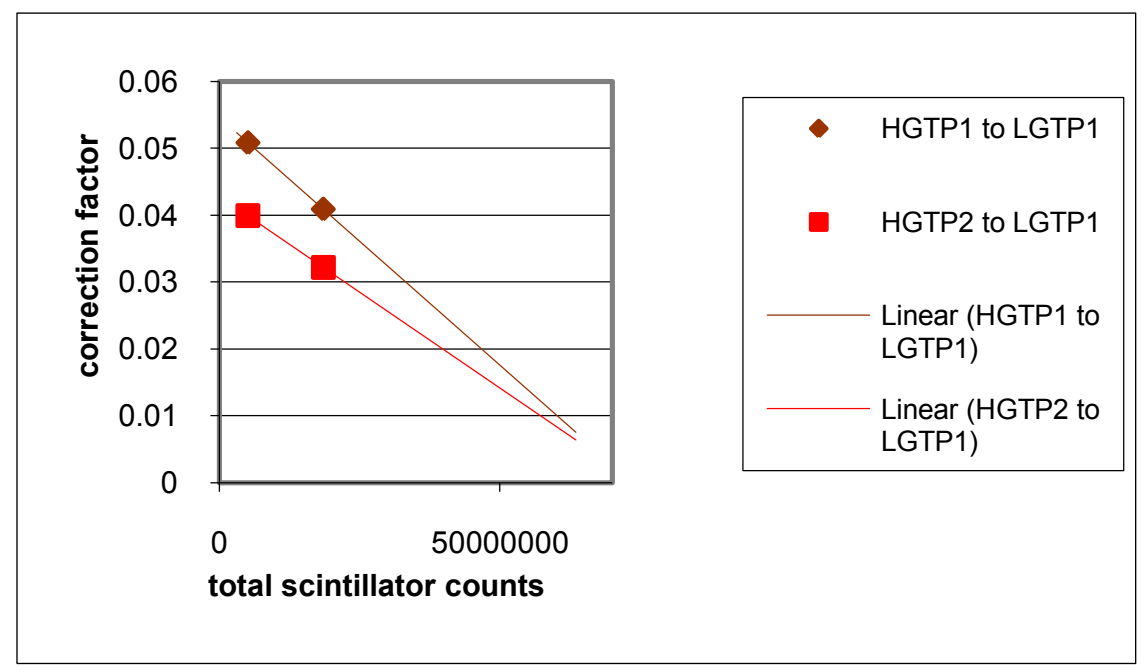

Fig. 4: Determination of the correction factors for temperature profiles 1 and 2 at high photomultiplier gain by means of a linear fit.

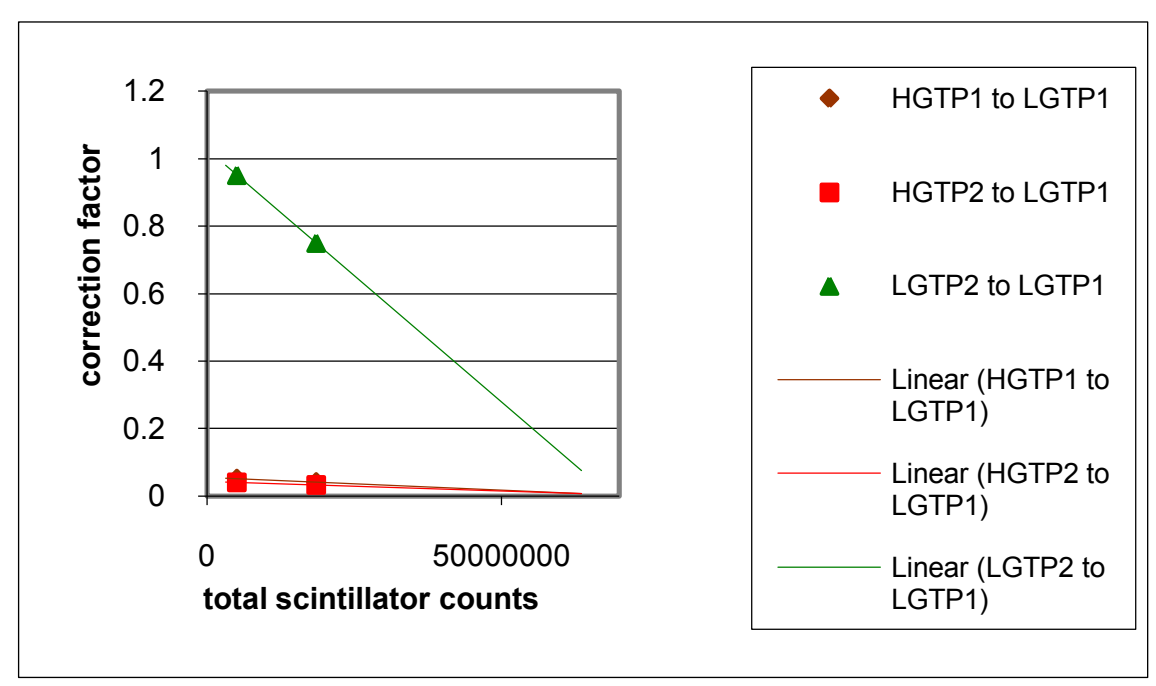

Fig. 5: Determination of the correction factor for temperature profile 2 at low photomultiplier gain by means of a linear fit. The curves for both temperature profiles at high photomultiplier gain (as in Fig. 4) are also shown.

In Fig. 6, the particle fluences obtained by using these correction factors are summarized and compared to the results obtained with the diamond sensor. Especially the normalized measurement obtained with the Harshaw reader is in good agreement with the diamond response. Taking the larger error into account, also the point at lower fluence obtained with the Alnor reader corresponds to an overall linear response of the sensors used.

Since only a limited number of TLD irradiations was possible during the validation experiment of the diamond sensor, the Alnor reader seemed to be appropriate for the lowest hadron fluence [4]. The additional uncertainty was taken into account by assigning a larger error to this single measurement, as shown in fig. 6.

| starter readout parameters | target readout parameters | correction factor $c_{xyz}$ |
|---|---|---|
| high gain, temperature profile 1 | low gain, temperature profile 1 | $c_{hi1}$ = -7E-10$x$ + 0.0546 |
| high gain, temperature profile 2 | low gain, temperature profile 1 | $c_{hi2}$ = -6E-10$x$ + 0.0428 |
| low gain, temperature profile 2 | low gain, temperature profile 1 | $c_{lo2}$ = -1E-08$x$ + 1.0272 |

Table 1: Correction factors as a function of the obtained dose. With these factors any readout regime can be calculated back to the one at low photomultiplier gain and temperature profile 1.

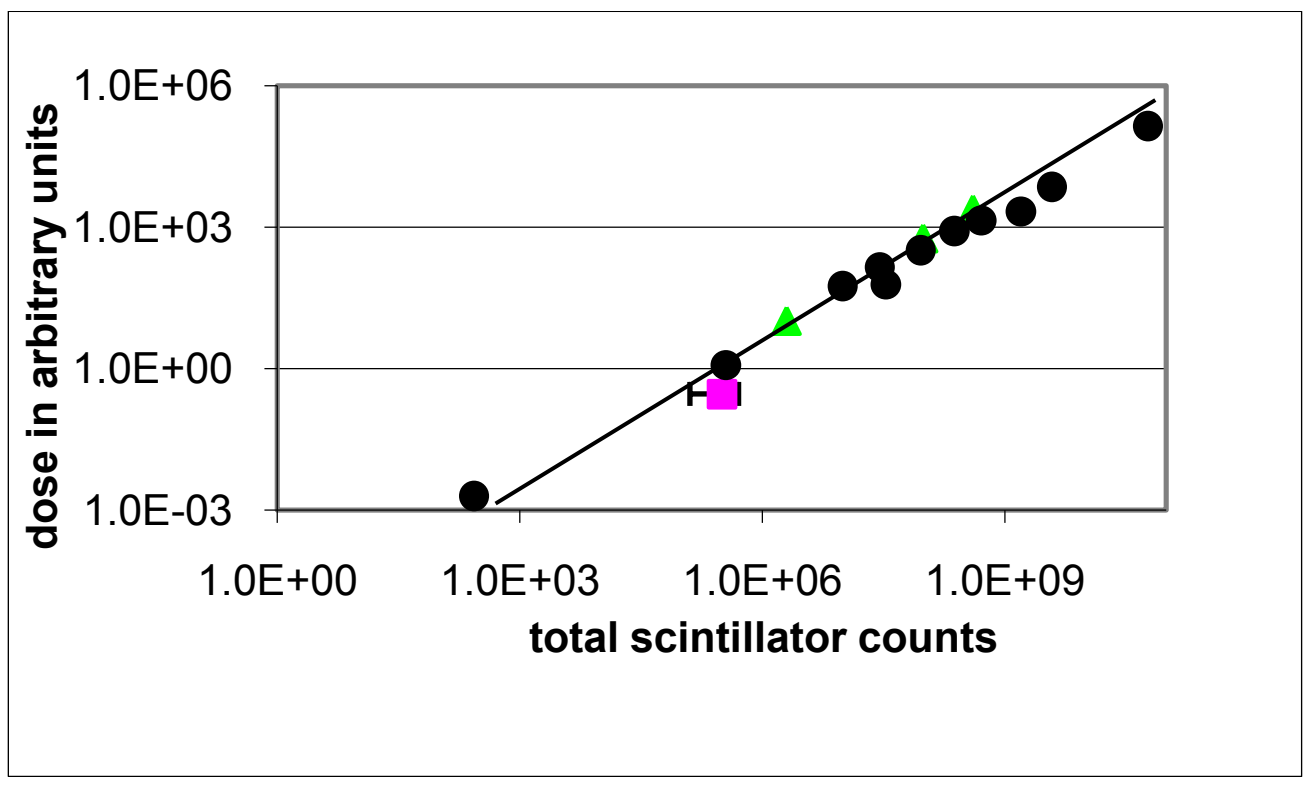

Fig. 6: Response of the fluence measurement with the Harshaw 5500 reader (green triangles) and the ALNOR Dosacus reader (purple square) after correction. The response of the diamond sensor is shown in black. The least-square fit that is shown is based on the TLD response only.

In the framework of this experiment, no analysis of potential radiation damage to the TLD could be performed. However, given the linearity of the sensor response over four orders of magnitude in dose, there is no hint on such radiation damage within the error of the measurement so far. The consistent shape of the beam profile as discussed in section 3 could be understood as an additional confirmation. Nevertheless, the possibility of radiation damage in TLD under hadron irradiation should be subject to further study.

## 3. Measurement of the beam profile

As an independent check on consistency of the fluence results obtained with the thermoluminescence detectors, a profile of the hadron beam from the accelerator was obtained from the analysis of a 7 × 7 matrix of TLD, using the Harshaw 5500 reader. The result is shown in Fig. 7, it is consistent with a beam profile as it is expected to be delivered by CERN's Proton Synchrotron.

Since this work intended to add dosimetric information to the calibration of diamond sensors, the linear response of that sensor over a wider range of particle fluxes is shown in Fig. 8.

## Summary

Apart from the fact that the diamond response to particle fluences in the considered range is linear within the error of the fluence measurement, it can be concluded that thermoluminescence dosimeters are suitable for measuring high-energy and high-fluence hadron fields. Nevertheless, for their reliable application, for example in similar fields in the LHC tunnel and experimental caverns, a more detailed calibration and analysis of their response to such fields, including potential radiation damage, should be done. Prior to using them for this application, the best readout device together with its appropriate settings needs to be determined.

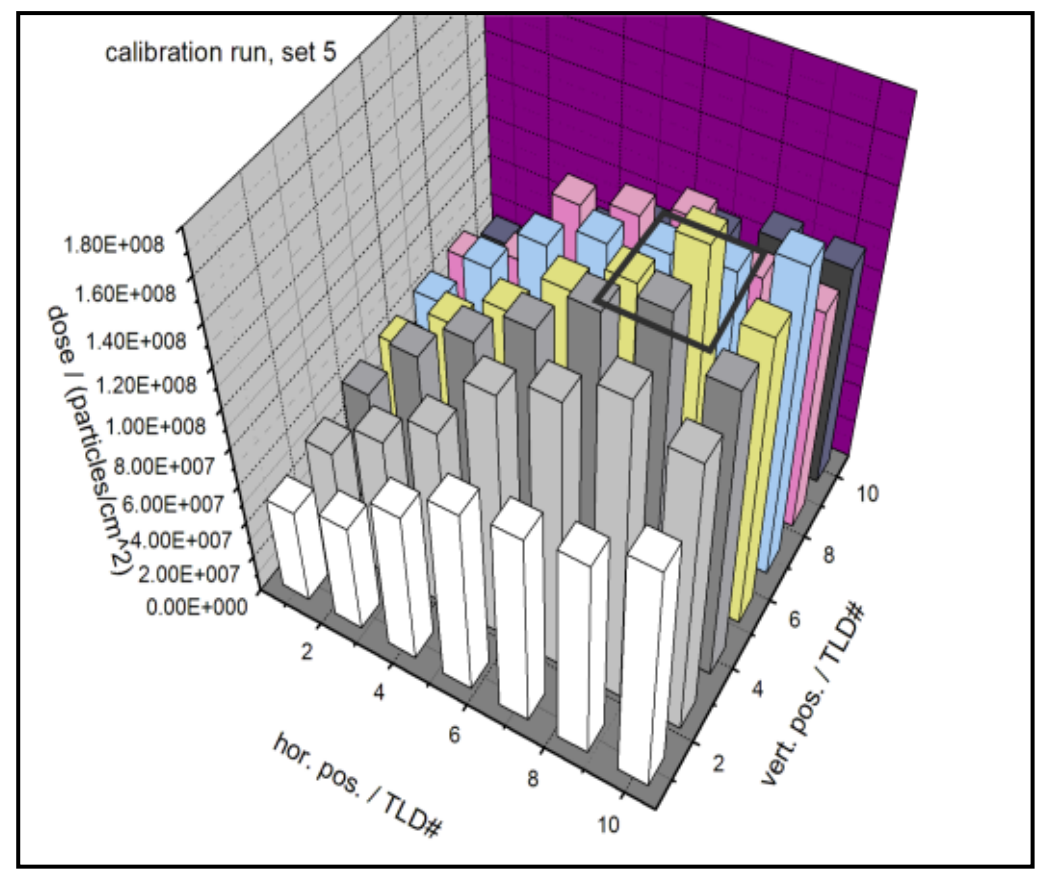

Fig. 7: Beam profile measured by a 7 x 7 matrix of thermoluminescence sensors, calibrated in absolute particle fluence. The black square is the surface of the scintillator used for particle counting.

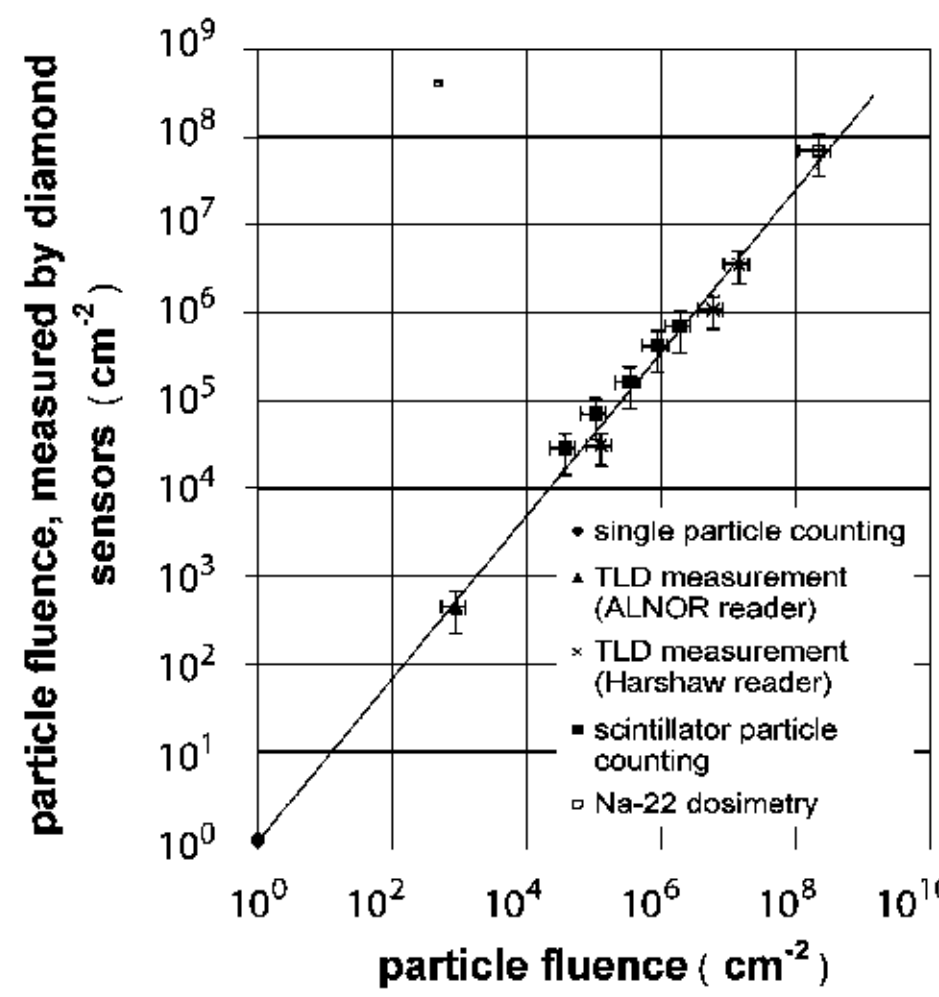

Fig. 8: Linear response of CVD diamond as a particle detector [2].

## Acknowledgments

The authors would like to thank R. Steerenberg for the development and operation of the fast extraction proton beam at CERN's Proton Synchrotron accelerator, and M. Glaser and F. Ravotti for their assistance with the aluminum activation measurements.

This work was supported by the German Ministry of Education and Research under grant number 05 HP6 PE1.